\documentclass[aps,floats,showpacs]{revtex4}

\input epsf.tex
\epsfclipon

\newcommand{\nc}{\newcommand}
\nc{\beq}{\begin{equation}}
\nc{\eeq}{\end{equation}}
\nc{\beqa}{\begin{eqnarray}}
\nc{\eeqa}{\end{eqnarray}}
\nc{\lra}{\leftrightarrow}

\nc{\sss}{\scriptscriptstyle}
{\nc{\lsim}{\mbox{\raisebox{-.6ex}{~$\stackrel{<}{\sim}$~}}}
{\nc{\gsim}{\mbox{\raisebox{-.6ex}{~$\stackrel{>}{\sim}$~}}}

\begin{document}

\title{Problems with Time-Varying Extra Dimensions\\ or ``Cardassian Expansion''\\ 
as Alternatives to Dark Energy}
\author{James M. Cline$^{\dagger,*}$ and J\'er\'emie Vinet$^*$}
\affiliation{$^\dagger$ Theory Division, CERN, CH-1211, Geneva 23, Switzerland\\
$^*$ Physics Department, McGill University, Montr\'eal, Qu\'ebec, Canada H3A 2T8}

\date{20 November 2002}

\begin{abstract}{It has recently been proposed that the Universe might be accelerating 
as a consequence of extra dimensions with time varying size.  We show that although these 
scenarios can lead to acceleration, they run into serious difficulty when taking into 
account limits on the time variation of the four dimensional Newton's constant.  On the 
other hand, models of ``Cardassian'' expansion based on extra dimensions which have been 
constructed so far violate the weak energy condition for the bulk stress energy, for 
parameters that give an accelerating universe.}
\end{abstract}


\maketitle

\section{Introduction}
In recent years, the physics community has witnessed a spectacular revival in interest 
for extra dimensions.  This revival was sparked by the realization that extra dimensions 
need not be small or compactified to agree with current experimental and observational 
constraints \cite{ADD}-\cite{RSII}.  The fact that they might lie within the reach of 
upcoming experiments while at the same time providing an elegant solution to the hierarchy 
problem has led to intense study of a number of models containing large and/or warped 
extra dimensions. 

Another subject which lately has attracted much attention is the observation that the 
Universe is accelerating \cite{SN1a}.  A number of hypotheses have been put forward in 
trying to explain this most unexpected result \cite{quintessence}-\cite{accmoffat}.  
Not surprisingly, given the current interest in branes and extra dimensions, many models 
have appeared which make use of these very ideas to obtain an accelerating Universe 
\cite{branes}-\cite{guhwang}.  

Here, we will be concerned with the mechanisms proposed in \cite{cardassian} and 
\cite{guhwang}, where the acceleration is due to the presence of extra dimensions.  
In \cite{cardassian} the essential new ingredient is some kind of bulk stress energy which 
changes the form of the Friedmann equations at late times, whereas in \cite{guhwang} the 
acceleration comes from time-variation of the size of the extra dimension. We will examine 
whether these ideas can lead to acceleration while complying with well-known physical 
constraints, namely on the time variation of the four dimensional Newton's constant or on 
the possible equation of state of the new form of stress energy. 

The plan of this note is as follows.  In section 2, we will give a brief review of the 
idea presented in \cite{guhwang}.  In section 3, we will confront this idea with experimental 
and observational constraints on the constancy over time of the gravitational force, showing 
that it runs into serious difficulty.  In section 4, we will analyze the relationship between 
this model and Brans-Dicke theory.  In section 5, we explore whether dropping the assumption 
that the extra dimensions are isotropic can alleviate some of the problems.  In section 6 we 
show that the model which has been proposed for Cardassian acceleration violates the weak energy 
condition in the bulk. We discuss our results and conclude in section 7.


\section{Acceleration From Time Variation of Extra Dimensions}

The idea starts with a Universe with our usual four spacetime dimensions, supplemented by $n$ 
compact dimensions with time varying size.  The metric will then have the form \cite{guhwang}
\beqa
ds^2 = dt^2-a^2(t) \left(\frac{dr_a^2}{1-k_ar_a^2}+r_a^2d\Omega_a^2\right) 
-b^2(t)\left(\frac{dr_b^2}{1-k_br_b^2}+r_b^2d\Omega_b^2\right) 
\eeqa
where $k_b = -1,0,+1$ characterizes the extra dimensions' spatial curvature.
This is to be used in the action
\beqa
\label{isoaction}
S&=&\int d^{4+n}x \sqrt{-\bar g}\left[-\frac{1}{16\pi\bar G}{\bar {\cal R}}+\bar{\cal L}_m\right] 
\eeqa
For a perfect fluid, with stress-energy tensor given by
\beqa
T^M_N = {\rm diag}(\bar \rho,-\bar p_a,-\bar p_a,-\bar p_a,-\bar p_b,-\bar p_b,....)
\eeqa
the Einstein equations will lead to 
\beqa
\label{Hsq}
\left(\frac{\dot a}{a}\right)^2 &=& \frac{8\pi \bar G}{3}\bar\rho - \frac{k_a}{a^2} 
-n\frac{\dot a}{a}\frac{\dot b}{b} 
- \frac{n(n-1)}{6}\left[\left(\frac{\dot b}{b} \right)^2+ \frac{k_b}{b^2}\right]\\
\label{addot}
\frac{\ddot a}{a} &=& -\frac{4\pi \bar G}{3}\left(\bar \rho + 3 \bar p_a\right) 
-\frac{n}{2}\frac{\dot a}{a}\frac{\dot b}{b} 
- \frac{n(n-1)}{6}\left[\left(\frac{\dot b}{b} \right)^2+\frac{k_b}{b^2} \right] 
- \frac{n}{2}\frac{\ddot b}{b}\\
\label{sum}
\frac{\ddot a}{a} + \left(\frac{\dot a}{a}\right)^2 &=& -\frac{8\pi \bar G}{3}\bar p_b 
-\frac{k_a}{a^2} -(n-1)\frac{\dot a}{a}
\frac{\dot b}{b} - \frac{(n-1)(n-2)}{6}\left[\left(\frac{\dot b}{b}\right)^2
+ \frac{k_b}{b^2} \right] - 
\frac{(n-1)}{3}\frac{\ddot b}{b}
\eeqa
Here, $\bar G, \bar \rho, \bar p_a$ and $\bar p_b$ refer to the $(4+n)$-dimensional quantities.   
The corresponding $4$-dimensional quantities will be written as $G_N,\rho,p_a,p_b$ where 
$G_N = \bar G/b^n$ and $[\rho,p_a,p_b] = b^n\times[ \bar \rho,\bar p_a,\bar p_b]$.  
Notice in particular the new terms in eq. (\ref{addot}) which depend on the extra dimensions.  
We see that there is the potential for a new source of acceleration beyond those of standard 
cosmology.  Let us now consider how constraints on the time dependence of Newton's constant 
constrain this possibility.

\section{Confrontation with Experimental Constraints}

It is well known that in models with extra dimensions, the effective strength of gravity 
is related to the volume of the extra  dimensions.  This dependence can be negligible in 
warped geometries \cite{RSI,RSII} but not in the type of factorizable geometry considered here.  
Indeed, in such a case, the effective $4$-D Newton's constant $G_N$ will be inversely proportional 
to the total  volume of the extra dimensions (see, e.g. \cite{ADD}).  Consequently, any variation 
in the extra dimensions' volume will show up as a variation of $G_N$,
\beqa
G_N\sim b^{-n}\Rightarrow \frac{\dot G_N}{G_N} = -n\frac{\dot b}{b}.
\eeqa

There are tight constraints on $\dot G_N/G_N$ from a number of experimental and observational 
considerations (see \cite{Gvar} for a thorough review).  The most generous upper bounds currently 
give, roughly, $\left|\dot G_N/G_N\right| < 3\times 10^{-19}s^{-1} $.  Combining this with the 
accepted value for the current Hubble rate \cite{hubble} $\frac{\dot a_0}{a_0}\approx 2,
3\times 10^{-18}s^{-1}$ leads to 
\beqa
\label{constraint}
\left|\frac{\dot b_0}{b_0}\right| \lsim \frac{1}{10n}\frac{\dot a_0}{a_0}. 
\eeqa
Given this bound, we can see immediately by looking at (\ref{addot}) that in the absence of 
curvature in the extra dimensions, only the term involving $\ddot b$ is capable of providing a 
significant positive contribution to the acceleration.  However, this term must then be of the same 
order as $\left(\frac{\dot a_0}{a_0}\right)^2$, a situation which appears somewhat unnatural given 
the fact that we demand that the first derivative of $b$ be much smaller. This might not be so 
unnatural if the extra dimensions were oscillating, since then we could accidentally be living at 
a time when $(\dot b/b)^2 \ll \ddot b/b$. However, for oscillations whose period is a significant 
fraction of the age of the Universe, to account for acceleration that was present at a redshift 
of $z=1$, the effective four dimensional theory would have to contain a nearly massless 
($m\sim H$) Brans-Dicke scalar field (the field $b$) with gravitational strength couplings 
to Standard Model particles.  If its mass were larger than $H$, it would oscillate on 
time scales $m^{-1}$ much shorter than the age of the universe. Such a light and
strongly coupled particle is clearly ruled out by experimental constraints
on Brans-Dicke-like theories, as we will discuss in more detail below.

Let us now look at what conditions will lead to acceleration, given (\ref{Hsq})-(\ref{sum}).  
Using (\ref{sum}) to eliminate the second derivative of $b$ from (\ref{addot}), and 
(\ref{Hsq}) to eliminate $\rho$,  we find that the conditions for getting a positive value 
of $\ddot a/a$ are
\beqa
\label{condition1}
\frac{\dot b}{b} > \frac{\dot a}{a}+\sqrt{\frac{(n+1)(n+2)}{n(n-1)}\left(\frac{\dot
a}{a}\right)^2+2\frac{2n+1}{n(n-1)}\frac{k_a}{a^2}-\frac{k_b}{b^2}
-\frac{16\pi G_N}{n(n-1)}\left((n-1)p_a-np_b\right)}\\
\label{condition}
\frac{\dot b}{b} < \frac{\dot a}{a}-\sqrt{\frac{(n+1)(n+2)}{n(n-1)}\left(\frac{\dot a}{a}\right)^2
+2\frac{2n+1}{n(n-1)}
\frac{k_a}{a^2}-\frac{k_b}{b^2}-\frac{16\pi G_N}{n(n-1)}\left((n-1)p_a-np_b\right)}.
\eeqa
Comparing with the constraint (\ref{constraint}), we can immediately rule out the first 
inequality (\ref{condition1}).

Since we are concerned with acceleration in our current era of matter domination, 
we will now set $p_a$ to zero.  Furthermore, measurements of the CMB imply that our three 
large spatial dimensions are flat, or have negligibly small curvature so that it is appropriate 
to set $k_a = 0$.  Demanding that (\ref{condition}) not conflict with (\ref{constraint}), we find 
the following:
\beqa
\label{const2}
{16\pi G_N}p_b-\frac{k_b}{b_0^2}(n-1) < \left(\frac{\dot a_0}{a_0}\right)^2
\left[\left(1+\frac{1}{10n}\right)^2(n-1)
-\frac{(n+2)(n+1)}{n}\right]
\eeqa
The right hand side is negative for all $n>0$, so this condition cannot be 
satisfied in the case studied in \cite{guhwang}, where $k_b=p_b=0$.

Having shown that the simple flat and pressureless option is excluded, we
now examine the more general cases where there can be pressure and/or
curvature in the extra dimensions. If $k_b$ is nonzero, then it should
play the role of the dark energy,
\beqa
-{n(n-1)\over 6 H^2 } {k_b\over b^2} = \Omega_\Lambda \sim 0.7
\eeqa
In no case do we expect $k_b > 0$, since this would strongly contradict
the latest estimates of $\Omega_\Lambda$ from CMB data \cite{wmap}. 
Therefore the inequality (\ref{const2}) only becomes harder to satisfy if we allow
for nonzero $k_b$.

Since the case $p_b=0$, $k_b = \pm 1$ is unacceptable, suppose instead that the 
extra dimensions are flat, i.e. $k_b= 0$.  Then (\ref{const2}) requires that $p_b < 0$.  
The only obvious way to do this is through the introduction of some extra matter component
to the energy density which would provide negative pressure along the extra dimensions, and 
most likely along the ordinary dimensions as well.  But this is precisely what the model was 
trying to avoid in the first place, by proposing the kinematics of the extra dimensions as 
the sole origin of the acceleration.  In light of this fact, it is hard to argue that such a 
model supplemented with some new form of stress energy  represents an improvement over other 
proposed explanations for dark energy.

\section{Relation to Brans-Dicke Theory}

We now analyze the relationship between the model presented above and Brans-Dicke theory \cite{BD}.  
Brans-Dicke (BD) gravity is a modification of general relativity where a scalar field couples to 
the Ricci scalar, thus yielding a theory in which the gravitational ``constant" will be time 
dependent.  The action is given in the simplest case by \footnote{Note that here as in the extra 
dimensional theory, $\bar G$ refers to the bare gravitational coupling. 
The effective gravitational constant will thus be $G_N = \bar G/\phi$.} 
\beqa
S = \int d^4x \sqrt{-g} \left(\frac{1}{16\pi \bar G} \left[-\phi{\cal R} 
+ \frac{\omega}{\phi}\partial_{\mu}\phi\partial^{\mu}\phi \right]
+{\cal L}_m\right),
\eeqa
where we assume that ${\cal L}_m$ refers to a perfect fluid.  We will generalize this slightly 
by adding a potential for the BD scalar
\beqa
\label{BDpotaction}S = \int d^4x \sqrt{-g}\left(\frac{1}{16\pi\bar G}\left[-\phi {\cal R} 
+ \frac{\omega}{\phi}\partial_{\mu}\phi\partial^{\mu}\phi-V(\phi)\right]+{\cal L}_m\right).
\eeqa
Models of this type have been the subject of a number of papers (see, e.g., \cite{BDgeneralizations}).  
We will now show explicitly the correspondence between the models obtained from (\ref{isoaction}) 
and (\ref{BDpotaction}).  The equations of motion for BD theory 
with a potential and a typical FRW metric $ds^2 = dt^2 - a(t)^2 dx_i^2$ are 
\beqa
\label{BDHsq}
\left(\frac{\dot a}{a}\right)^2 &=& \frac{8\pi \bar G}{3\phi} \rho 
+ \frac{\omega}{6}\left(\frac{\dot \phi}{\phi}\right)^2
-\frac{\dot a}{a}\frac{\dot \phi}{\phi}+\frac{V(\phi)}{6\phi}\\
\label{BDaddot}
\frac{\ddot a}{a} &=& -\frac{4\pi \bar G}{3\phi}\left(\rho + 3 p\right) 
- \frac{\omega}{3}\left(\frac{\dot \phi}{\phi}\right)^2
-\frac{1}{2}\frac{\dot a}{a}\frac{\dot \phi}{\phi}-\frac{1}{2}\frac{\ddot \phi}{\phi} 
+\frac{V(\phi)}{6\phi}\\
\label{BDsum}
(2\omega +3)\left(\frac{1}{3}\frac{\ddot \phi}{\phi}
+\frac{\dot a}{a}\frac{\dot \phi}{\phi}\right) &=& 
\frac{8\pi \bar G}{3\phi}\left(\rho-3p\right)+\frac{1}{3}\left(2\frac{V(\phi)}{\phi}
-\frac{\partial V(\phi)}{\partial \phi}\right).
\eeqa
Comparing these with (\ref{Hsq})-(\ref{sum}), we see that the following
choice (which was also obtained in \cite{radosc}) leads to complete equivalence between 
both sets of equations, as long as the extra dimensional pressure term $\bar p_b$ 
vanishes: 
\beqa
\label{equivs}
\phi&\equiv& b^n\nonumber\\
\omega&\equiv& \frac{1}{n}-1\nonumber\\
V(\phi) &\equiv& -n(n-1)k_b b^{n-2}\nonumber\\ 
&=& -n(n-1)k_b \phi^{1-2/n} 
\eeqa
 We can now apply all known constraints on the parameters of BD theory to the model 
of \cite{guhwang}.  We note here two important points.  The first is that for a 
vanishing potential, experimental limits demand that $\omega > 1500$ \cite{radosc}.  
Clearly, this is not possible in the present context.  Furthermore, the potential given 
here is non-trivial for $n>2$, in which case its minimum is located at $\phi = 0$, for 
which $G_N$ diverges.    

What can we say about the case $\bar p_b \neq 0$?  From the point of view of BD theory, 
we can look at this in the following way.  BD theory is explicitly constructed to limit the 
effect of the scalar field to inducing time dependence in the strength of gravity.   This means 
that matter will not be directly affected by the presence of the scalar, so that conservation 
of energy will be given by the  standard  
\beqa
\label{Econs}
\dot \rho = -3\frac{\dot a}{a}(\rho+p)\nonumber\\
\Rightarrow \rho \sim a^{-3(1+\omega)}.
\eeqa 
If we now look at the conservation of energy equation in the extra dimensional theory, it 
will be given by
\beqa
\label{modEcons}
\dot {\bar\rho} = -3\frac{\dot a}{a}(\bar\rho+\bar p_a)-n\frac{\dot b}{b}(\bar\rho +\bar p_b).
\eeqa
Writing $\bar p_a = \omega_a \bar\rho$ and $\bar p_b = \omega_b \bar\rho$, and assuming 
$\omega_a$ and $\omega_b$ to be constant, 
we have
\beqa
\bar\rho \sim a^{-3(1+\omega_a)}b^{-n(1+\omega_b)}\\
\Rightarrow \rho \sim a^{-3(1+\omega_a)}b^{-n\omega_b}.
\eeqa
We can see that if $\bar p_b = 0$ then these will reduce to the standard form (\ref{Econs}).  
However, if $\bar p_b \neq 0$, conservation of energy will be modified.  Indeed, in order 
for a scalar-tensor theory to behave as the extra dimensional model does in the case where 
$\bar p_b\neq 0$, the energy density would need to scale as (using (\ref{equivs}))
\beqa
\rho &\sim& a^{-3(1+\omega_a)}\phi^{-\omega_b}\nonumber\\
\Rightarrow \dot \rho &=& -3H(1+\omega_a)\rho-\omega_b\frac{\dot\phi}{\phi}\rho
\eeqa
leading to a theory qualitatively different from BD gravity.

\section{Anisotropic Extra Dimensions}

In the model we have examined above, it was assumed that all the extra dimensions had 
the same scale factor.  We will now drop this assumption and see whether some of the difficulties 
we have pointed out for isotropic extra dimensions can be relieved.  Such models, dubbed 
{\it multidimensional cosmology} have already been extensively studied (see \cite{multi,multi2} 
and references therein), but, as far as we know, not with a view toward obtaining cosmic 
acceleration (however see ``Note Added'' at end).

A particularly interesting aspect of such models is the fact that in the corresponding 
multi-scalar-tensor theory, only one scalar (corresponding to the total volume of the extra 
dimensions) couples to matter.  This means that constraints on the time-variation of Newton's 
constant will only apply to this one field.  From the point of view of the extra dimensional 
theory, one might therefore hope that as long as the total volume remains approximately 
constant, the variations of the individual dimensions can be large enough to have the 
desired effect on the dynamics of the Universe.  

The metric for this model will be given by \footnote{Following the discussion in section 3, 
we will assume from the start that the extra 
dimensions are flat.}
\beqa
\label{metric}
ds^2 = dt^2 - a^2(t)dx^2 -\sum_i c_i^2(t) d\theta_i^2
\eeqa
Where $i$ runs from $1$ to $n$, the number of extra dimensions.  The stress-energy tensor will 
be that of a 4+n dimensional perfect fluid
\beqa
T^m_n &=& {\rm diag}(  \bar \rho,-\bar p,-\bar p,-\bar p,-\bar p^{\theta_1},\dots, 
-\bar p^{\theta_n}).
\eeqa

The Einstein equations then lead to the following:
\beqa
\label{Hsqmod}
\left(\frac{\dot a}{a}\right)^2&=&\frac{8\pi \bar G}{3}\bar \rho
-\frac{\dot a}{a}\sum_i \frac{\dot c_{i}}{c_{i}}-\frac{1}{6}
\left(\sum_{i}\frac{\dot c_{i}}{c_{i}}\right)^2+\frac{1}{6}\sum_{i}
\left(\frac{\dot c_{i}}{c_{i}}\right)^2\\
\label{accmod}
\frac{\ddot a}{a} &=& \frac{-4\pi \bar G}{3}\left[\bar \rho+ 3 \bar p\right]
-\frac{1}{2}\frac{\dot a}{a}\sum_i\frac{\dot c_{i}}{c_{i}}
-\frac{1}{6}\left(\sum_{i}\frac{\dot c_{i}}{c_{i}}\right)^2
-\frac{1}{2}\frac{d}{dt}\sum_i\frac{\dot c_{i}}{c_{i}} 
-\frac{1}{3}\sum_{i}\left(\frac{\dot c_{i}}{c_{i}}\right)^2
\eeqa
\beqa
\label{55mod}
\frac{8\pi \bar G}{3}p^{\theta_k}-\frac{1}{3}\frac{d}{dt}\left(\frac{\dot c_{k}}{c_{k}}\right)
-\frac{\dot c_{k}}{c_{k}}
\left(\frac{\dot a}{a}+\frac{1}{3}\sum_i\frac{\dot c_{i}}{c_{i}}\right)&=&
-\frac{4\pi \bar G}{3}\left(\bar \rho-3 \bar p\right)+
\frac{1}{6}\left(\sum_i\frac{\dot c_{i}}{c_{i}}\right)^2 \nonumber\\&&
+\frac{1}{2}\frac{\dot a}{a}\sum_i\frac{\dot c_{i}}{c_{i}}
+\frac{1}{6}\frac{d}{dt}\sum_i\frac{\dot c_{i}}{c_{i}}.
\eeqa
Note that there will be one equation of the form (\ref{55mod}) for each of the $n$ 
extra dimensions, whereas for isotropic extra dimensions, these equations were degenerate.

The same reasoning we employed in the previous section allows us to state that 
\beqa
\label{anisotropicconstraint}
\left| \sum_i \frac{\dot c_{i}}{c_{i}}\right| \lsim  \frac{1}{10}\frac{\dot a_0}{a_0}.
\eeqa
Notice, however, that due to the independence of the $\frac{\dot c_{i}}{c_{i}}$'s, this in 
no way constrains the $\sum_i \left(\frac{\dot c_{i}}{c_{i}}\right)^2$ terms appearing in 
the Friedmann equations.  This means that it is in principle possible that the sum of the 
squares term in (\ref{Hsqmod}) accounts for the missing energy density, while the 
$\frac{d}{dt}\sum_i\frac{\dot c_{i}}{c_{i}}$ term in (\ref{accmod}) accounts for the 
acceleration.  As in the isotropic case though, it seems unnatural to ask that the 
time derivative of a small quantity be large.  Furthermore, we have not yet taken 
(\ref{55mod}) into account.

Adding the $n$ equations (\ref{55mod}) leads to 
\beqa
\frac{8\pi \bar G}{3}\sum_i \bar p^{\theta_i}-\frac{1}{3}\frac{d}{dt}\sum_i 
\left(\frac{\dot c_{i}}{c_{i}}\right)
-\sum_i \frac{\dot c_{i}}{c_{i}}\left(\frac{\dot a}{a}
+\frac{1}{3}\sum_i\frac{\dot c_{i}}{c_{i}}\right)&=&
-n\frac{4\pi \bar G}{3}\left(\bar \rho-3 \bar p\right)
+\frac{n}{6}\left(\sum_i\frac{\dot c_{i}}{c_{i}}\right)^2\nonumber\\
&&+\frac{n}{2}\frac{\dot a}{a}\sum_i\frac{\dot c_{i}}{c_{i}}
+\frac{n}{6}\frac{d}{dt}\sum_i\frac{\dot c_{i}}{c_{i}}
\eeqa
which allows us to write
\beqa
\frac{\ddot a}{a}&=& -4\pi \bar G\left[\frac{\bar \rho}{3}
\left(1+\frac{3n}{2+n}\right)+\bar p\left(1-\frac{3n}{2+n}\right)
+\frac{2}{2+n}\sum_i \bar p^{\theta_i}\right]\nonumber\\
&&+\frac{\dot a}{a}\sum_i\frac{\dot c_i}{c_i} 
+ \frac{1}{3}\left(\sum_i \frac{\dot c_i}{c_i}\right)^2
-\frac{1}{3}\sum_i\left(\frac{\dot c_i}{c_i}\right)^2.
\eeqa
Using (\ref{anisotropicconstraint}) and the fact that $\bar p=0$ in our current 
matter-dominated epoch, we see that in order to obtain a positive value for 
$\ddot a/a$, we must have $\sum_i \bar p^{\theta_i} <0$, as was the case for 
isotropic extra dimensions.


\section{``Cardassian'' expansion from extra dimensions}

In the previous sections we considered time-varying extra dimensions as a source of 
cosmic acceleration.  Also in the context of extra dimensional effects, it was 
recently proposed \cite{cardassian,gf} that the current era of acceleration arises 
from modifying the Friedmann equation by adding a term proportional to $\rho^n$:
\beq
\label{cardass}
	H^2 = {8\pi\over 3} G(\rho + C\rho^n)
\eeq
The authors have dubbed this kind of expansion ``Cardassian.''  The new term
dominates at late times if $n < 1$, and if $n<2/3$, it gives rise to a positive
acceleration.  
This kind of behavior is qualitatively very different from the standard 
braneworld result which has $n=2$ \cite{braneworld}, because it implies a modification 
of gravity at very low energy scales rather than very high ones.  Actually 
eq.\ (\ref{cardass}) is a bit misleading, in giving the impression that $\rho$ is the 
sole source of expansion.   All of the particle physics models which have been proposed 
to give the modified Friedmann equation introduce an exotic new source of energy which 
contributes significantly to the expansion; the effects of the new matter are merely 
parametrized in terms of the conventional matter energy density $\rho$.  In the present 
discussion we will treat only the extra-dimensional model of Cardassian expansion on which 
\cite{cardassian} is based. ({In the alternative model of \cite{gf}, a new form of dark 
matter with an exotic confining force is hypothesized, in which the interaction energy 
redshifts more slowly than cold dark matter, and this gives rise to the $\rho^\alpha$ 
term.  We note that there exists no realistic microscopic model for obtaining the kind of 
dark matter needed by \cite{gf}, since they require a new confining force with a confinement 
scale larger than the present horizon, and this force must be mediated by seemingly 
implausible objects with spatial dimensionality greater than 1.})

The Cardassian model in question is based on work of ref.\ \cite{chung-freese}, which
showed that if one parametrizes the Hubble rate in terms of the brane  energy density 
$\rho$ (as though the bulk contribution to the energy density was hidden), then  an 
arbitrary power $n$ can be obtained for the Friedmann equation, $H^ 2 \sim \rho^n$.  
This conclusion was reached by writing down a candidate solution for the 5D Einstein 
equations and Israel junction conditions, and then seeing what form of the bulk stress
energy tensor would be required to make this indeed a solution to the bulk equations.
Specifically, they consider a 5D metric of the form
\beq
	ds^2 = e^{2\nu(t,r)} dt^2 - e^{2\alpha(t,r)} d{\bf x}^2 - e^{2\beta(t,r)} d{r}^2 
\eeq
where the extra dimension is bounded by branes at $r=0$ and at $r=l$.  The metric functions
are given by 
\beq
\label{card-soln}
	\beta(t,r) = \nu(t,r) = c r t^{-2/n};\qquad \alpha(t,r) = -\frac12\beta(t,r) 
+ \frac{2}{3n}\ln t
\eeq
with $c = \frac13\kappa_5^2 \mu \left(\frac{2}{3n}\right)^{2/n}$, where $\mu$ is 
a dimensionful constant such that the Friedmann equation is exactly $H^2=(\rho/\mu)^n$.  
(No solution has been proposed which gives what one really wants, namely a transition 
from $H^2\sim\rho$ to $H^2\sim \rho^n$.)

The size of the extra dimension is time-dependent in the solution 
given by (\ref{card-soln}); this may be incompatible with stabilizing its size, 
as is usually required for a braneworld model to be compatible with fifth force constraints.  
However $\beta$ asymptotically approaches a fixed value at large times, so it may be 
possible that such a solution is consistent with having a stabilized radion, as well 
as satisfying constraints on the time-dependence of the gravitational force. In the present 
model, Newton's constant is proportional to 
\beq
	G_N \propto \int_0^l e^{\beta} dr = 
\left.G_N\right|_{t\to\infty}{t^{2/n}\over c l}\left(e^{clt^{-2/n}} - 1\right)
\eeq
which approaches a constant as $t\to\infty$.  There seems to be sufficient freedom in 
the choice of parameters to insure the relative constancy of $G_N$, even since the era 
of big bang nucleosynthesis.

However, the model runs into difficulties when we examine the equation of state of the 
bulk stress energy, in particular the $T_5^{\ 5}$ component: we find that the weak energy 
condition is violated, with the pressure component $|T_5^{\ 5}|$ being larger in magnitude 
than $T_0^{\ 0}$.  Specifically,
\beqa
\kappa_5^2 T_0^{\ 0} &=& 
\frac{{e^{-2\,{c}\,{t}^{-q}r}}}{3t^2}\,{{ \left( {q}^{2}-\frac{9}{4}\,{t}^{-2
\,q}{{c}}^{2}{r}^{2}{q}^{2}-9\,{t}^{-2\,q+2}{{c}}^{2}
 \right) }} \\
\kappa_5^2 T_5^{\ 5} &=& 
	-\frac{{e^{-2\,{c}\,{t}^{-q}r}}}{3t^2}\,{\left( \frac{9}{2}\,{t}^{-q}{c
}\,rq-9\,{t}^{-2\,q}{{c}}^{2}{r}^{2}{q}^{2}-\frac{9}{4}\,{t}^{-2\,q+2}{{
c}}^{2}-2\,{q}^{2}-\frac{9}{2}\,{t}^{-q}{c}\,r{q}^{2}+3\,q \right) }\nonumber
\eeqa
where $q=2/n$.  In the large-time limit, the ratio of $T_5^{\ 5}$ to  $T_0^{\ 0}$ is
\beq
	{T_5^{\ 5}\over T_0^{\ 0}} = 2 - \frac32 n
\eeq
Interestingly, this quantity starts to violate the weak energy condition just for 
the values of $n$ where acceleration begins, namely $n< 2/3$!   We remind the reader 
that such a violation implies unphysical behavior; in field theory a negative kinetic 
energy is implied, corresponding to a wrong-sign kinetic term which spoils unitarity.

\section{Conclusion}

The idea put forth in \cite{guhwang} that the Universe's current accelerated expansion 
might have purely geometrical origins initially sounds quite attractive.  Indeed, it is 
easy to show that general multidimensional models can lead to our 3-space accelerating 
without invoking new matter components with exotic equations of state.  However, as we 
have argued, the most simple of such models are completely equivalent to a  class of 
Brans-Dicke theories which are ruled out by experiment, either because the BD parameter 
$\omega$ is too small, or because the scalar field potential has a minimum for which 
Newton's constant diverges.  The only way to make these models compatible with experimental 
constraints while still leading to acceleration is to invoke non-zero pressure along the 
extra dimensions.  This, however, means that we are no longer dealing with ``normal'' 
matter, thus spoiling the model's initial motivation.  This is true for anisotropic as 
well as for isotropic extra dimensions$.$  From the point of view of scalar-tensor theory, 
we have shown that this non-vanishing pressure along the extra dimensions corresponds to 
modifying conservation of energy such that the theory differs from standard BD gravity. 

As far as ``Cardassian'' expansion is concerned, we have focused on a particular proposal 
which makes use of extra dimensions in order to obtain a modified Friedmann equation.  
It appears that it is possible to tune the parameters in such a way that Newton's constant 
tends toward a constant value fast enough to avoid any conflict with experimental evidence.  
However, we have shown that upon closer inspection, the bulk stress energy tensor behaves 
in a way which leads to a serious difficulty.  Indeed, for exactly the values of the parameter 
$n$ which make acceleration possible, it violates the weak energy condition.  That being said, 
it would be very interesting if someone could construct a plausible microphysical model which 
leads to the Cardassian hypothesis without violating fundamental principles like causality or 
unitarity.

There are alternative ways \cite{branes} of using extra dimensions to explain acceleration, 
relying on the intrinsic curvature of branes to achieve the desired effect.  Other works 
(e.g. \cite{radosc,evmod,binetruy}) have also investigated the cosmological implications 
of the evolution of the radion.  Such models evade the constraints we have pointed out here, 
and as such are currently on sounder footing from a theoretical standpoint than the ones analyzed 
in this paper.

It is clear that extra dimensions lead to a number of interesting effects which might provide 
solutions to a number of outstanding questions in fundamental physics.  In light of our results 
however, it is also clear that much work needs to be done on the models we have analyzed here 
if they are to  be considered as prime candidates in resolving the mystery of cosmological 
acceleration.

\section{Acknowledgements}
J.C.\ and J.V.\ are supported in part by grants from Canada's National Sciences and Engineering 
Research Council.

\section{Note Added} After completion of this work, we were informed by M.\ Pietroni of an 
anisotropic model \cite{pietroni} which gives a more negative equation of state ($w=-1/3$) than 
does ordinary dark matter ($w=0$).  This result is consistent with our observation in section V 
that negative values for the extra-dimensional pressure components are needed to get cosmic 
acceleration.  The model of ref.\ \cite{pietroni} has positive pressures, and it does not have 
acceleration, since $w=-1/3$ is the borderline value between deceleration and acceleration. We 
thank M.\ Pietroni for bringing this work to our attention.

We also wish to thank U.\ Guenther and A.\ Zhuk for pointing out \cite{multi2}, and O.\ Bertolami 
for pointing out \cite{BBS} to us.

\end{document}